\documentclass{emulateapj}
\usepackage{CJKutf8}
\usepackage{graphicx,amsmath,amsthm,ulem,color}
\usepackage{natbib}
\usepackage{color}
\usepackage{amssymb}

\newcommand{\qp}{\textcolor{black}}
\newcommand{\qr}{\textcolor{black}}
\newcommand{\be}{\begin{equation}}
\newcommand{\ee}{\end{equation}}
\newcommand{\dt}{{\partial t}}

\newcommand{\del}{\nabla}
\usepackage{epstopdf}
\def\refnew#1{(\ref{#1})}

\newcommand{\blue}{\textcolor{black}}
\newcommand{\gsim}{\lower.7ex\hbox{$\;\stackrel{\textstyle>}{\sim}\;$}}
\newcommand{\lsim}{\lower.7ex\hbox{$\;\stackrel{\textstyle<}{\sim}\;$}}

\begin{document}

\title{Ohmic Dissipation in Mini-Neptunes}
\author{Bonan Pu$^1$ \& Diana Valencia$^{2,3}$}
\affil{$^1$ Department of Astronomy \& Space Sciences, Cornell University, Ithaca, NY 14853, USA \\
$^2$ Centre for Planetary Sciences, Department of Physical $\&$ Environmental Sciences, University of Toronto, Toronto, Canada \\
$^3$ Department of Astronomy \& Astrophysics, University of Toronto, Toronto, Canada\\}

\begin{abstract}

In the presence of a magnetic field and weakly ionizing winds, ohmic dissipation is expected to take place in the envelopes of Jovian and lower-mass planets alike.  While the process has been investigated on the former, there have been no studies done on mini-Neptunes so far.  From structure and thermal evolution models, we determine that the required energy deposition for halting the contraction of mini-Neptunes increases with planetary mass and envelope fraction. 
\blue{Scaled to the insolation power, the ohmic heating needed is small $\sim10^{-5}$ -- orders of magnitude lower than for exo-Jupiters $\sim 10^{-2}$.  Conversely, from solving the magnetic induction equation,  we find that ohmic energy is dissipated more readily for lower-mass planets and those with larger envelope fractions.  Combining these two trends, we find that ohmic dissipation in hot mini-Neptunes is strong enough to inflate their radii  ($\sim 10^{15}$ W for $T_{eq}=1400K$).  The implication is that the radii of hot mini-Neptunes may be attributed in part to ohmic heating. Thus, there is a trade-off between ohmic dissipation and H/He content for hot mini-Neptunes, adding a new degeneracy for the interpretation of the composition of such planets. In addition, ohmic dissipation would make mini-Neptunes more vulnerable to atmospheric evaporation.}
 \end{abstract}

\section{Introduction}

As the research field of exoplanets has progressed with more discoveries uncovering trends in the data, there are several outstanding questions that need to be answered.  In terms of structure and composition there are arguably two main unresolved issues: For the mini-Neptunes it is important to determine what the composition of their small envelopes is, as this will carry information as to where and perhaps how they formed. And for exo-Jupiters we need to explain why so many of them are inflated beyond a composition made out of pure H/He, the so called radius-anomaly. 

One common thread to addressing in part both these questions, is to study how ohmic dissipation affects the structure and thermal evolution of a planet with an atmosphere.  While this mechanism has been heavily studied for the purposes of explaining the radius anomaly of exo-Jupiters \citep{BS10,Perna_2010_hotJ, magscale, BS11, wu_lithwick, xuhuang, rogers_1, rogers_2}, and to constrain the wind structure on Jupiter and Saturn \citep{Liu}, here we focus on assessing how ohmic heating affects our inference of the composition of mini-Neptunes.

Mini-Neptunes are planets that have a low enough mass that they do not have substantial atmospheres (i.e. $\lesssim 15 M_\oplus$) but are large enough that they cannot be made out of solid material entirely, a threshold that varies as a function of mass. For a given mass and radius measurement, these planets are larger than the terrestrial threshold radius, so that we deduce they have an envelope. However, it is not possible to determine if the composition of the small envelope is made mostly out of H/He or water, or both, on the basis of mass and radius measurements alone because of how inherently degenerate the problem is \citep{valencia_ternary, Rogers:2010, valencia_2013}. Here, we explore how ohmic dissipation in mini-Neptunes affects their structure and radius, and thus compositional inference. 

In the context of exo-Jupiters, it is well established that a significant number of close-in transiting hot Jupiters \blue{ exhibit radii that are larger} than expected from standard models of giant planet evolution (see review \citet{fortney_nettelmann}). A variety of mechanisms have been invoked to explain this radius anomaly and can be divided into three categories (following \citet{weiss}): incident flux-driven mechanisms (that include kinetic heating \citep{showman_guillot} and ohmic dissipation \citep{BS10, Perna_2010_hotJ},  tidal mechanisms \citep{tides} and delayed contraction mechanisms (that include enhanced opacities \citep{burrows} and supression of heat transport \citep{CB_heat} such as layered convection \citep{Leconte_layered}.  For more details, see the review by \citet{Baraffe_review}. Recently \citet{Tremblin:Tadvection} has proposed yet a different mechanism by which atmospheric 2D circulation in tidally-locked planets leads to hotter interior than 1D models predict via the advection of potential temperature and thus a larger planetary radius.  
%\blue{Works driven by data of exoJupiters \citep{Guillot:radanom, Laughlin:radanom, Enoch:radanom, weiss}  have tried to shed a light on which mechanisms are the most important by fitting radii to different characteristics such as mass, temperature, host star metallicity, insolation, etc.}

While it is very possible that a combination of mechanisms is responsible for the inflation of exo-Jupiters, there is evidence that a dominant effect causing inflation is absent at warm and cold effective temperatures. Using internal structure and evolution models on 14 warm exoplanets, \citet{miller_inflation} suggested that planets receiving less stellar irradiation than $\sim2\times10^{8}$ erg/sec/cm$^{2}$ (or $10^{5}$ W/m$^{2}$) equivalent to an equilibrium temperature of 1400 K, did not appear inflated (i.e. planets were smaller than a pure H/He planet). In addition, observational work on 115 Kepler candidates by \citet{Demory_inflation} revealed a trend between planetary radius and incident radiation above $\sim2\times10^{8}$ erg/sec/m$^{2}$ only, showing consistency in the results. \blue{This trend was also observed by \cite{weiss} to only hold true for massive planets ($M \ge 150 M_{\oplus}$) and suggested a weak scaling of planetary radius and incident flux. \cite{Laughlin:radanom} analyzed the radius anomalies (difference between the observed and predicted radii) of hot exoJupiters and found a scaling with effective temperature that was too steep (coefficient in power law of $\alpha \simeq 1.4$) to be attributed to kinetic heating ($\alpha_{kin} = 0.67$) but consistent with ohmic dissipation ($\alpha_{ohm} \simeq  2.4$). On the other hand, \cite{Enoch:radanom} found a shallower radius anomaly scaling (of $\alpha \simeq 0.84$) when metallicity and semi-major axis of the planet were taken into account.} With this evidence, ohmic dissipation stands out as a viable mechanism to explain inflation in hot Jupiters. Ohmic dissipation is both a mechanism that produces inflation (by increasing the interior entropy of the planet from dissipation within the convective interior), and delays contraction (by moving the convective-radiative boundary deeper in the interior from dissipation in the atmosphere). 

Energy dissipation due to ohmic effects is expected to be present in planets that have weakly-ionized winds and a magnetic field, regardless of the planet's mass. The ions in the atmospheric wind moving across the planetary magnetic field induce electrical currents that can dissipate ohmic energy deep in the interior by removing kinetic energy from the atmospheric wind \citep{Liu, BS10}. The primary source of electrons in the atmosphere comes from alkali metals (e.g. Na, Al, and K) that are thermally ionized, leading to a strong dependence of ohmic dissipation with atmospheric temperature. This is consistent with the dependency of hot-Jupiters' radius anomaly with temperature.  In addition, this process is self limiting as highly ionized winds start freezing with the planetary magnetic field, slowing down the wind through Lorentz drag \citep{Perna_2010_hotJ, magscale} imposing an upper limit in temperature for ohmic heating to take place.

%While ohmic heating is arguably the leading theory in explaining the radius anomaly, a few studies have suggested that the effect is not large enough \citep {xuhuang,rogers_2}.

Supported by evidence that ohmic dissipation may take place in hot-Jupiters, we aim our efforts in understanding the effect of ohmic dissipation in mini-Neptunes for two reasons. 
To assess how it will affect our inference of envelope composition given that if large enough, ohmic dissipation may mascarade as more H/He content adding to the sources of compositional degeneracy for these planets. And second, by looking at another region of the parameter space, we hope to shed light on the debate of how ohmic dissipation operates on planets.  We start with a brief history of how ohmic dissipation has been treated so far as to put into context our modeling and assumptions.

\subsection{Previous Studies}

In 2010, \citeauthor{BS10}, inspired by \citet{Liu}, proposed that ohmic dissipation (a new mechanism at that time) could inflate hot Jupiters. They solved the magnetic induction equation in steady
state in a weakly ionized atmosphere in the presence of a planetary dipole magnetic field provided a prescribed wind speed profile. They used an exponential approximation to the conductivity profile that enabled them to solve the equations analytically in the atmosphere, and numerically in the interior for the induced current, and ohmic power as a function of radius up to a constant. This constant is adjusted so that the amount of irradiation that gets converted into ohmic energy (i.e. efficiency factor) stays fixed at the few percent level. They ignored the effect of Lorentz drag and looked at static interior models for three exoJupiters as examples. 

The same year, following up on the work by \citet{Perna_2010_Mdrag} on magnetic drag,  \citeauthor{Perna_2010_hotJ} independently suggested ohmic dissipation to explain the radius anomaly solving for the wind profile via atmospheric circulation models of a tidally locked planet, calculating the ohmic power generated via the azimuthal component of the induced current envisioning a slightly different geometry. While \citet{BS10} had the radial currents looping throughout the interior depositing
the ohmic energy deep within the planet, \citet{Perna_2010_hotJ} solved for the meridional currents (ignoring radial component) depositing the energy at pressures of several tens of bars, still deep enough to halt contraction during the thermal evolution of a planet \citep{showman_guillot}. Another difference is that \citet{Perna_2010_hotJ} include the Lorentz drag in the atmospheric wind \qr{(via a linear friction that relaxes the wind towards zero over a specified drag time)}, and show that it reduces ohmic heating by about a factor of 5 for HD209548b, not enough to preclude ohmic dissipation as a mechanism for exoJupiters inflation provided they have a strong magnetic field ($\sim10$ G). 

The next year \citet{BS11} coupled an internal structure and evolution planetary model to their ohmic dissipation results to track which exoJupiters would be inflated given their age. They could prescribe
ohmic dissipation as a function of radius given a pressure-temperature profile for a (coreless) planet and calculate consistently its thermal evolution. They also included a \qp{linear (Rayleigh)} prescription for Lorentz drag. Their most notable result shows that around $1200-1800$ K, there is an increase in radius inflation due to ionization for exoJupiters. Cool planets do not have enough ionization to generate ohmic dissipation, and very hot planets (above $\sim1800$ K) have somewhat saturated ionization levels and deeper convective regions that hinder deep energy deposition. They were able to expain all the inflated planets at the time. In fact, their nominal efficiency ($\sim1$\%) and coreless model over estimated the radii of the low-mass Jupiters $\left(\lesssim0.7M_{J}\right)$. By extension, the radii of these planets could be easily explained
either by invoking a large core $\left(20\,M_{E}\right)$ or a smaller efficiency. 

\citet{wu_lithwick} revisited the subject and with a different model for internal structure and conductivity prescription obtained similar results. Ohmic dissipation can inflate exo-Jupiters provided an efficiency
of a few percent ($\sim3\%$ ) is prescribed. To obtain the desired efficiency they adjust the wind velocity. Furthermore, they changed the depth of the wind zone and noticed that deeper atmospheric winds
can inflate planets more easily (the efficiency invoked for inflation may be lower). 

The same year, \citet{magscale} proposed simple scalings by doing an order of magnitude balance between the acceleration of the zonal flow, the presure gradients (related to thermal response of the atmosphere)
and the magnetic drag in the atmosphere. The study finds that magnetic drag can eventually slow down the winds to the point of imparing ohmic dissipation. The behaviour of normalized ohmic dissipation to irradiation increases with temperature up to levels of $\sim1\%$ and then decreases. The location for the peak depends on the magnitude of the magnetic field and for nominal values of 3-30 G, it happens between 1300-1900K. However, these values depend strongly on the atmospheric response to irradiation which are poorly known (e.g. opacities, thermal inversions, etc). Importantly, \citet{magscale} found that the magnetic Reynolds number is larger than one for hot planets (above 1500K for magnetic fields in the order of 3-10 G), indicating a degree of coupling between the flow and the magnetic field, with the possibility of a poloidal induced field in the atmosphere in addition to that of the planet. 

\citet{xuhuang} implemented a different approach to previous studies by looking at the wind zone and interior of the planet separately. They solved the induction equation in the interior and connected the
solution to the wind zone by setting boundary conditions for the temperature and toroidal magnetic induced field (or radial induced current) disregarding initially that the two are connected. Subsequently, they relate the temperature below the windzone and the toroidal magnetic field via the scalings from \citet{magscale}, and find that for the allowed combinations of temperature and toroidal field, ohmic dissipation is too low and cannot account for the radius inflation of most exoJupiters. For all other quantities fixed, they find that the ohmic power decreases with mass $\sim R^{4}/M$, and once they include the effects of variable
conductivity and feedback of ohmic dissipation on the structure, the relationship changes to $P_{ohm}\sim R^{2.4}/M$. 

Later on, \citet{rogers_2} implemented an MHD code to investigate the radius inflation of HD209458b. They too find that the magnetic Reynolds number for realistic flows exceeds unity and as a consequence they find that the Lorentz force is very efficient in slowing down the winds leading to low values of ohmic dissipation. However, even without ohmic drag, their wind profile is considerably slower than results coming from conventional GCM atmospheric studies. They obtain values near $0.1$ km/s at the 1 bar level, one order of magnitude less than conventional studies \citep{Heng_rev:2015, Showman_rev:2010}.  Given that ohmic dissipation scales as the velocity square, this may understimate the ohmic dissipation calculated by two orders of magnitude.  In addition, they mention two other assumptions that can influence their results.  First, their choice of boundary conditions that can affect the calculation of the radial current and thus, the heating rate profile, and second, the fact that conductivity is dependent on the reference state temperature (instead of the actual local temperature). In addition, like all atmospheric models, the geometry is limited to the outer most shell that limits the study of deep deposition by radial currents.

Follow up work by \citet{rogers_1} extended this study to include effects of different reference temperatures, day-night temperature differences, magnetic field strenghts, as well as viscous, thermal and magnetic diffusivities.  They find that adding a magnetic field to a hydrodynamic treatment produces more complicated effects in the flow (oscillatory, stationary and westward mean flows) than those captured by pure drag (that cause reduced easward flow), and that ohmic dissipation seems too low (about an order of magnitude) to explain the radius inflation of Hot Jupiters, but would affect more lower-mass planets. The caveats they mention to their results are several including that  their implementation of an anelastic treament precludes the resolution of fast wind speeds commonly seen in the atmospheres of Hot Jupiters (which may explain why the obtain such very slow winds). Other assumptions include the fact that their model extends to 200 bars while ohmic dissipation is most efficient when it happens at depth where perhaps convective motions are important, the need to have a coupled evolutionary models with MHD,  as well as their implementation of a constant magnetic diffusivity.

Most recently, \cite{Sari_1, Sari_2} developed an analytical model for the inflation of hot Jupiters due to deep energy deposition. They assumed a power-law opacity and a simple equation of state including both thermal and an electron degeneracy pressure. They found that heat deposition can generate an exterior convective region which serves to delay the cooling of the planet, provided that the heat source is sufficiently deep. The critical quantity is the fraction of energy deposited compared to the equilibrium luminosity times the depth at which the energy is deposited. They find that the estimated a critical efficiency of $\sim 5\%$ is required to reproduce the observed hot Jupiter radii; in comparison, they estimate the peak ohmic dissipation efficiency to be $\sim 0.3$, reached at a temperature of $1500$ K. Such efficiencies can explain inflated radii up to $R \approx 1.6 R_J$, but fail to explain some of the more extremely inflated planets.

% The critical quantity is the ratio $L_{\mathrm{dep}} \tau_{\mathrm{dep}} / L_{eq}$, where $L_{\mathrm{dep}}$ and $L_{eq}$ are the deposited luminosity and equilibrium luminosity respectively, and $\tau_{\mathrm{dep}}$ is the depth of deposition in terms of optical depth. In their model, the inflated planet radius scales as $\Delta R \propto \left(1 + \frac{L_{\mathrm{dep}}\tau_{\mathrm{dep}}}{L_{eq}}\right)^{\beta(1-\beta)/(4 - \beta b)}$, where $\beta \approx 0.35$ is a parameter determined by the opacity scaling and the equation of state. Applying this result to ohmic dissipation, \cite{Sari_2} estimated a critical efficiency of $\epsilon \approx 5\%$ is required to reproduce the observed hot Jupiter radii; in comparison, they estimate the peak ohmic dissipation efficiency to be $\epsilon_{\mathrm{max}} \approx 0.3$, reached at a temperature of $1500$ K. Such efficiencies can explain inflated radii up to $R \approx 1.6 R_J$, but fail to explain some of the more extremely inflated planets. 

In sum, while it is well accepted that ohmic dissipation should be present in highly irradiated planets there is still no consensus on the magnitude of the magnitude of this effect and its efficiency in inflating hot Jupiters, while a few of these studies point to the fact that lower-mass planets would be more amenable to radius inflation.

\subsection{This work}
 In contrast to previous studies, which focused on planets with Jovian masses, in this study we direct our attention to mini-Neptune planets. These planets differ from exo-Jupiters in two key aspects: Firstly, they feature lower surface gravities due to their lower masses, so that their envelopes are less gravitationally bound. For example, the gravity of Corot-24b, a Neptune-like planet, is $g = 4.2$ m/s$^2$ \citep{corot-24b} and six times lighter than Jupiter's gravity $g = 24.8$ m/s$^2$ or 2 times lighter than HD209458b with $g = 9.4$ m/s$^2$ . Secondly, unlike Jupiters, which are entirely gaseous in their composition, mini-Neptunes are dominated by a rocky core, with the gaseous envelope accounting for less than $10 \%$ of the total mass in most cases \citep{valencia_2013} .  As a result, the internal heat flux may be largely affected by the rocky core, and the geometry of the current flow is a thinner shell (assuming currents do not penetrate into the rocky interior, (see below), as opposed to a sphere for the case of Jupiter.

We assume an internal structure that is  a rocky interior overlain by a small envelope (up to $10\%$ by mass) which describes most of the mini-Neptunes found  \citep{valencia_2013} and as a first step do not consider ices in the interior based on the fact that we are mostly interested in planets that are close to their star. It is unclear if mini-Neptunes are formed in situ with an absence of water/ices \citep{Chiang_SE_form}, or formed beyond the snow line with subsequent migration \citep{Hansen_SE_form}. Thus, we consider the in-situ scenario as a first step in understanding how ohmic dissipation would influence such type of planets. 

To study ohmic dissipation in mini-Neptunes we follow a two step approach. We first model the interior structure and thermal evolution of these planets with various degrees of energy deposition to obtain envelope profiles. With these results, we construct the conductivity profile and solve the induction equation given a prescribed wind profile and obtain the current distribution and ohmic dissipation power. For self-consistent solutions, we feed the calculated ohmic dissipation back into the original planet interior structure models, and iterate until the derived ohmic dissipation strength matches the heat deposited in the model.  The ohmic dissipation feedback on the planet's internal structure profile is parametrized as a single point heat source deep in the planet's interior, and taken to be constant in time. This approximation was found to be consistent with time evolution models of mini-Neptunes, as models with significant ohmic dissipation will reach radiative equilibrium at fixed final constant dissipation value. As a first step, in this study we do not account for the effect of Lorentz drag on the atmospheric flow.

We find that ohmic dissipation can indeed play a role in inflating the radii of hot mini-Neptunes, analogous to the case of hot Jupiters. 

This manuscript is organized as follows: in section 2 we discuss the details of our modelling procedures,  in section 3, we present the results of our calculations and the implications on inferring the composition of mini-Neptunes, in section 4 we discuss implications, as well as the validity and limitations of our assumptions, and in section 5 we summarize our results. 

\section{Model}
For this study, we consider planets that have masses ranging from $2 M_{\oplus}$ to $16 M_{\oplus}$, and with gas envelope fractions $f \equiv M_{env}/M_{tot}$ ranging from $1 \%$ to $10 \%$. The composition of the gaseous envelope is assumed to be $X = 0.70$, $Y = 0.28$ and $Z = 0.02$, where $X$, $Y$ and $Z$ are the mass fractions of hydrogen, helium and metals respectively. Solar abundances are used for the relative abundances of metals within $Z$.  \qr{For the purposes of this study we ignore the possibility of the envelopes of mini-Neptunes having substantial water. }

We focus on planets with an equilibrium temperature $T_{eq}$ between $1300$ K and $1700$ K. We find that, for temperatures below $1300$ K, ohmic dissipation is insignificant due to lack of ionization in the atmosphere, whilst at temperatures above $1700$ K, for smaller planets ($M \le 9M_{\oplus}$) ohmic dissipation leads to planetary radius expansion ($dR/dt > 0$) in most cases when Lorentz drag is not taken into account. Therefore, such temperatures are at the upper limits of the regime of validity of this study.

\subsection{Interior structure modelling}
To calculate the evolution of the interior structure of mini-Neptunes, we use CEPAM, a code originally derived from the stellar evolution code CESAM but with additional physics relevant to planetary bodies incorporated \citep{CEPAM}. Through combining the standard atmospheric structure models with the realistic model for rocky objects described in \citet{valencia_06}, this model is capable of simulating differentiated planets with an Earth-like rocky core surrounded by a gaseous envelope composed of hydrogen, helium and/or water. The algorithm divides the rocky interior of the planet into radial shells corresponding to the iron core, mantle, and the gaseous envelope, and solves their structural evolution in each region, and joined at the boundaries by mass and pressure continuity \citep{valencia_2013}. It includes internal heat sources as radioactive heating in the rocky mantle in chondritic proportions. For the gaseous portion of the planet, the structure of the atmosphere is determined based on a semi-grey model described in \citet{guillot_atm}, and extends to an optical depth of $\tau = 10^6$.  Below such depths, the \qr{temperature} structure of the gaseous envelope is determined using the lesser of the radiative and adiabatic gradients. We assume that the planetary magnetic field is generated in the convective liquid iron core \citep{Driscoll_Olson}. 

In this work, we define the envelope as the entire portion of the planet that is comprised of gaseous H/He. The uppermost layers of the envelope, from the planet's exterior down to an optical depth of $\tau = 10^6$ is referred to as the 'atmosphere' and we call the interior rocky portion of the planet, containing both the mantle and the iron core, as the 'core'. The value of maximum optical depth of the atmosphere is rather arbitrary but not important as long as it is high enough to map into the interior envelope. The overall planet radius $R$ is taken to be the chord radius of the planet's envelope, defined in \citet{guillot_atm}.

In our interior models, initial entropies of planets were set to be sufficiently large as to ensure a 'hot start'. Planets that form at high temperatures go through an initial rapid cooling phase, during which the photosphere temperature cools down to $T_{ph} \sim T_{eq}$. \cite{Sari_1} estimated this timescale to be $t \sim 1$Myr at $T_{eq} = 1500$K (with numerical values adjusted to suit mini-Neptunes), after which planet temperatures converge to the same state regardless of their initial starting temperatures, provided that the initial temperature $T_0 \gsim T_{eq}$. In this work, we define a 'hot start' loosely as any initial temperature sufficiently high to reach the convergent state (typical values of $S_0 \sim 1 \times 10^{9}$ erg/K/g at age zero). We ran the structure and thermal evolution models for up to $10$ Gyr.

\begin{figure}
\includegraphics[width=0.45\textwidth,trim=0 0 0 0, clip]{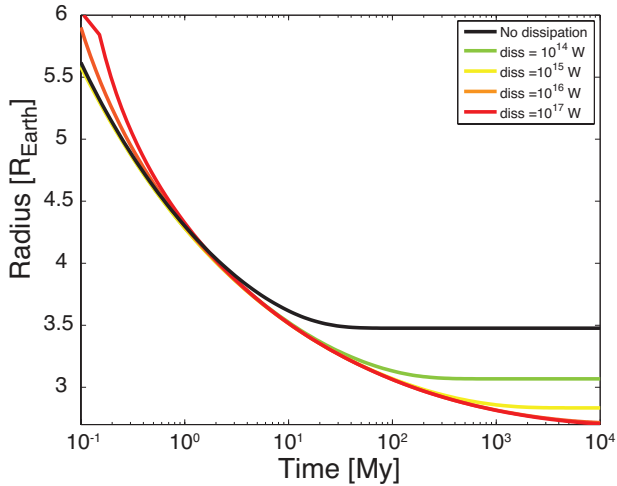}
\caption{Typical planet radius evolution for different amounts of ohmic dissipation in the planet's interior. This particular figure shows the cooling track of a $9 M_\oplus$- planet with 5\% H/He and $T_{eq} = 1500$K.}
\label{Revo_ohm}
\end{figure}

We consider cases with no additional internal heating (only radioactive), and models with ohmic dissipation. A typical radius evolution for a mini-Neptune contracting only under the irradiation of the host star is shown in Figure \ref{Revo_ohm} (black curve), the other curves have dissipation added.  The planet's envelope decreases in size as the planet undergoes Helmholtz contraction, until an equilibrium is reached; the exact time and the planet's final radius and temperature depends on the planet's properties and the presence of any other heat sources. For the cases without additional heating, most of the radius contraction happens before 1Gy. 

\begin{figure}
\includegraphics[width=0.45\textwidth,trim=0 0 0 0, clip]{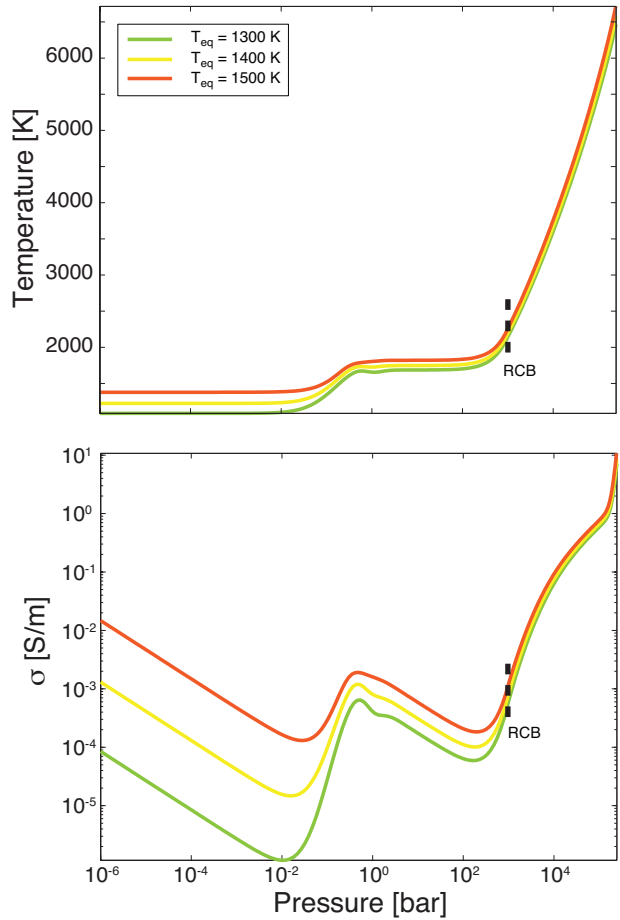}
\caption{Typical atmospheric and conductivity profile of a mini-Neptune. Calculations correspond to a $M = 9 M_{\oplus}$ planet with $f = 5\%$ for various equilibrium temperatures, taken as a snapshot after $1$ Gyr of evolution. Top: The temperature-pressure profile in the envelope is characterized by two isothermal layers, the outer one with a temperature $T \sim 0.8 T_{eq}$ and the inner one with a temperature of $T \sim 1.2 T_{eq}$. Bottom: The conductivity profile is characterized by a minimum and maximum values around the non-isothermal portion of the atmosphere. The radiative-convective boundary occurs at $\sim 1000$ bars and moves gradually inward during evolution as the planet cools.}
\label{TP_cond}
\end{figure}

Figure \ref{TP_cond} (top) shows a typical pressure-temperature profile for a H/He envelope of a mini-Neptune. The structure is defined by two nearly isothermal structures below a pressure of $\sim $ mbar, corresponding to a temperature of $T_{atm} \sim 0.8 T_{eq}$; and between pressures of $\sim 1$  to $\sim 10^{3}$  bars, with $T \simeq T_{rcb} \sim 1.2 T_{eq}$ corresponding temperatures, where $RCB$ stands for the radiative-convective boundary, with a steady increase within these two isothermal layers. Above $\sim 10^{3}$  bar, the convective region of the envelope begins and becomes adiabatic.  Generally, at the very early stages of the evolution the $RCB$ is very shallow ($< 100$ bars or even $<10$ bars). By 10 Myr the boundary is typically at 200 bars, and  by 100 Myr, unless the planet's contraction is significantly halted, the $RCB$ is between 1000 and 2000 bars (typically). If the planet cools to 10 Gyr without additional energy sources, the $RCB$ eventually reaches very deep pressures - $2 \times 10^4$ bars. The trend is that planets with thinner atmospheres (e.e. less H/He \%) have deeper $RCB$s, and more massive planets have shallower $RCB$s. The rate of $RCB$ becoming deeper is related to the rate of cooling. The $RCB$ no longer moves if the planet reaches equilibrium, and it also moves slower for planets at high equilibrium temperatures. The reason we care about the behaviour of the $RCB$ is because its size is connected to how much ohmic dissipation is available deep in the interior for inflation. 

\subsection{Conductivity profile}
As stated before, we assume a geometry where the currents do not penetrate the rocky interior and are confined to the thin envelope. It is unclear what the conductivity values for rocks would be under the interior conditions of mini-Neptunes ($T>5000$ K and $P = 10^5$ bars or $10$ GPa). The conductivity of rock minerals depends not only on temperature, but also sensitively on the presence of impurities controlled by factors such as the amount of water (hydrogen) and oxygen fugacity \citet{Karato_cond}. While there is experimental data on upper mantle materials, lower mantle materials such as perovskite and postperovskite, which would constitute the bulk of the interior of mini-Neptunes \citep{valencia_highP} is absent. Because of these uncertainties, as a first step we regard the rocky outer shell of the mini-Neptunes as an insulator, and confine the induced currents to the envelope. 

Having calculated the envelope structure ($P,T$), we compute the conductivity profile. Two sources could potentially contribute to electrical conductivity in the atmospheres of mini-Neptunes. At low pressures, solar irradiation causes partial thermal ionization of alkali metals, with elements K, Na, Li, Cs, and Fe being the dominant source of ions. Deeper in the gaseous envelope, the conductivity profile is dominated by the pressure-ionization of hydrogen, which dominates over ionization at pressures greater than $P \sim 10^5$ bar. 

The thermal ionization is described by the Saha equation:
\begin{equation}
\label{eq:saha}
\frac{n_{j}^{+}n_{e}}{n_{j} – n_{j}^{+}} = \left(\frac{m_e k_b T}{2 \pi \hbar^2}\right)^{3/2} \exp{(-I_j / k_b T)},
\end{equation}
where $n_j$ and $n_j^+$ are the total and positively ionized number densities of constituent $j$ respectively, $n_e$ is the number density of electrons, $m_e$ is the electron mass, $k_b$ is the Boltzmann constant,  $T$ is temperature, $\hbar$ is reduced Planck’s constant, and $I_j$ is the ionization potential of constituent $j$. For this study, we calculated the relevant contributions of the first 28 elements of the period table, from He through Ni, with hydrogen being treated separately.  The electrical conductivity of such a gas $\sigma_Z$ is given by:

\begin{equation}
\label{eq:sigmaz}
\sigma_{Z} = \frac{n_{e}}{n} \frac{e^2}{m_e \nu},
\end{equation}
where $n_e$ is the electron number density, $n$ is the total number density of particles, and $\nu$ is the collision frequency of electron-neutral collisions given by \citet{xuhuang}

\begin{equation}
\nu = 10^{-15} n \left(\frac{128k_BT}{9\pi m_e}\right)^{1/2} cm^3 s^{-1}.
\end{equation}

As the temperatures in question are usually much lower than the first ionization energy of even the most easily ionized element, K (with an ionization energy of 4.35 eV), it turns out appropriate to only consider the element K, which gives a conductivity \citep[for a derivation, see][appendix]{xuhuang}

\begin{equation}
\sigma_Z \approx 1.74 \times 10^3 \left(\frac{T}{1600K}\right)^{3/4} \left(\frac{p}{bar}\right)^{-1/2}  e^{-4.35eV /  k_B T}.
\end{equation}

Below pressures of $\sim 100$ bars, the above expression is accurate to within a few percent compared to our full implementation of the Saha equation. At higher temperatures and pressures, the above approximation breaks down as ionizations of other metals become significant. 

Deeper in the planet's interior, pressures are high enough to induce pressure ionization of hydrogen, and at such depths hydrogen becomes a semi-conductor. For the conductivity, we use the prescription from appendix A4 in \citet{xuhuang}):

\begin{equation}
\sigma_{X} = \sigma_0 \exp{\left(\frac{-E_g(\rho)}{k_B T}\right)},
\end{equation}

where $E_g = 20.3 - 64.7 \rho$ eV, $\rho$ is the density in mol cm$^{-3}$, and $\sigma_0 = 3.4 \times 10^{10}$ in SI units. The total conductivity is then calculated as

\begin{equation}
\sigma = \sigma_X + \sigma_Z.
\end{equation}

An example of the conductivity profiles for a $6 M_{\oplus}$ planet with $5 \%$ H/He content and different equilibrium temperatures is shown in Fig. \ref{TP_cond} (bottom).

%\begin{figure}
%\includegraphics[width=0.45\textwidth,trim=0 0 0 0, clip]{Figures/fig2_done.eps}
%\caption{An example conductivity profile for a $M = 6 M_{\oplus}$ planet with $f = 5\%$. The figure shows the conductivity as a function of pressure for various equilibrium temperatures, taken as a snapshot after $1$ Gyr of evolution. The radiative convective boundary is slightly at different pressures for the different temperatures but within the marking shown in the figure.}
%\label{conductivity}
%\end{figure}

\subsection{Wind Profile}
To solve the induction equation, we need the conductivity structure and wind profile. For planets on very short orbits (periods of days), aside from being subject to intense irradiation from their host stars, these tight orbits also cause the planets to be tidally locked, such that the same side of the planet always faces the host star. This tidal locking mechanism creates strong winds flowing from the day-side to the night-side in the upper atmosphere of these planets. In our models, it is this wind that produces the currents responsible for ohmic dissipation in the outer atmosphere. Studies of global circulations on hot Jupiters have shown that winds can reach $1$ to $2$ km/s at the 1 bar level on these planets. Similar studies by \citet{gj1214b, Kataria_GJ1214GCM} on the 6.5 $M_{\oplus}$ planet GJ 1214b showed that much less massive planets can attain very similar wind speeds as hot Jupiters. It turns out, in the absence of strong magnetic drag, the zonal wind speed due to temperature gradients only on tidally locked planets can be approximated by \citep{magscale}:

\begin{align}
v_{\phi} &\sim \sqrt{k_B \Delta T \Delta \log{P} / \mu} \\
&\sim 1.4 \text{km/s} \times \left(\frac{\Delta T}{500 K}\right)^{1/2}  \left(\frac{\mu}{2 \text{amu}}\right)^{1/2} (\ln{\Delta P})^{1/2},
\end{align}

 where $v_{\phi}$ is the wind velocity, $k_B$ is the Boltzmann constant, $\mu$ is the mean molecular mass, $\Delta T$ is the day-night temperature difference, and $\Delta \log{P}$ is the depth of the weather layer. For values typical of hot-Neptunes and inflated planets in general, the temperature difference is a substantial \qr{fraction} of $T_{eq}$, and this results in winds of $\sim 1 $ km/s. In this study, we assume the same maximal wind speed of $1$ km/s throughout our planet models.

\qr{Following the typical envelope structure of the two nearly isothermal layers we find a conductivity maxima and minima  (see Fig. \ref{TP_cond}). The minimum conductivity value happens at the base of the shallower isothermal layer ($\sim 10$ mbar) and the maximum at the top of the deeper isothermal layer and $\sim 1 $ bar.}

In our set-up, we assume that the base of the wind zone starts at $p = 10$ bars, and that the winds grow in intensity from this depth up until reaching the top of the lower, hotter isothermal sphere, and then stays constant throughout the atmosphere (see Fig. \ref{induction}). Above the base of the wind zone at the pressure of 10 bars, we assume the winds to take the expression \citep{BS10} :

\begin{equation}
v_\phi(r) = v_{max} \left(\frac{r - r_{10}}{r_{iso} - r_{10}}\right)^2 ,
\end{equation}

where $r_{iso}$ is defined to be the location of the conductivity maxima in the atmosphere, $r_{10}$ is the radius at $p = 10$ bars, and $r$ is the radius. Above $r > r_{iso}$, we assumed a constant wind speed of $v_{\phi}(r) = 1$ km/s. Note that the quantity $(r - r_{iso})$ is the same as $\delta$ used by \citet{BS10}.

\subsection{Ohmic Dissipation}

Ohmic theory has been worked out in detail by different groups (\citet{Liu,Perna_2010_hotJ, BS10, BS11, xuhuang, wu_lithwick}, among others). We summarize their key findings below.

For a current density $\mathbf{J}$ and conductivity $\sigma$, the volume power density of ohmic dissipation is given by:

\begin{equation}
\label{eq:ohmicdiss}
dP_{Ohm} = \frac{J^2}{\sigma} dV \sim 4 \pi r^2 \frac{J^2}{\sigma} dr .
\end{equation}

The current density is related to the planet's magnetic field $\mathbf{B}$ and the atmospheric wind velocity $\mathbf{v}$ through Ohm's law:
\begin{equation}
\label{eq:ohmslaw}
\mathbf{J} = \sigma \left(- \nabla \Phi +  \frac{\mathbf{v}}{c} \times \mathbf{B}  \right)    ,
\end{equation}

with $\Phi$ being the electric potential. The motion of the atmospheric wind under a fixed magnetic dipole field $\mathbf{B_{dip}}$ generates an induced magnetic toroidal field given by the induction equation:

\begin{equation}
\label{eq:induction}
\frac{\partial\mathbf{B}}{\dt} = \del \times (\mathbf{v} \times \mathbf{B}) - \del \times \lambda (\del \times \mathbf{B}).
\end{equation}

Under assumptions of steady-state and continuity, the solutions to the electric potential are then given by:
\begin{equation}
\del \cdot \sigma \del \Phi = \del \cdot \sigma (\mathbf{v} \times \mathbf{B_{dip}}).
\end{equation}

To calculate the distribution of currents inside the planet in detail, we consider a simple geometry with a zonal flow given by $\mathbf{v} = v(r) \sin{\theta} \hat{\phi}$. To solve the above equation in spherical coordinates, we can write the induced toroidal field as:
\begin{equation}
\mathbf{B_{ind}} = \frac{g(r)}{r} \sin{\theta}\cos{\theta}\hat{\phi},
\end{equation}

whose radial dependence satisfies

\begin{equation}
\label{eq:g}
\frac{-4\sigma M}{r} \frac{d}{dr}\left(\frac{v(r)}{r}\right) = \frac{d^2 g(r)}{dr^2} - l(l+1) \frac{g(r)}{r^2} - \frac{d \ln{\sigma}}{dr}\frac{dg(r)}{dr},
\end{equation}

where $M$ is the magnetic dipole moment of the planet and $l = 2$ is the index of the associated Legendre polynomial $P_l^1$. The current $\mathbf{J}$ is determined by Ampere's law $\mathbf{J} = (c / 4 \pi) \nabla \times \mathbf{B}$. In this study, we solve equation \refnew{eq:g} numerically to obtain the current distribution and induced magnetic fields, which we then used to evaluate the strength of ohmic dissipation.

Before the second-order equation \refnew{eq:g} can be solved, we must first pick two appropriate boundary conditions. \citet{BS10} made the physical argument that the conductivity drops several orders of magnitude over the transition between the two isothermal layers, and therefore the conductivity minima there could be treated as an insulating shell; they assumed that all currents would vanish at this point and integrated inward from the conductivity minimum. In a later study, \citet{BS11} fixed the boundary conditions to match a certain prescribed dissipation efficiency $\epsilon$. \citet{xuhuang} used a different approach and  solved the induction equation in the interior and connected the solution to the wind zone by setting boundary conditions for the temperature and toroidal magnetic induced field (or radial induced current) disregarding initially that the two are connected. Subsequently they relate the temperature below the windzone and the toroidal magnetic field via the scalings from\citet{magscale}. 

For this work we pick our boundary conditions by letting the radial current $j_r$ vanish at a pressure of $p = 10^{-6}$ bar (e.g. insulating shell). We also run models using the prescription from \citet{BS10}. We find that the two choices for boundary conditions agree well below temperatures of $T_{eq} \lesssim 1400$ K. Beyond this temperature, solar insolation heats up the outer isothermal layer enough that the conductivity drop across the layers is only a factor of $\sim 50$ less, so that the approximation of treating the conductivity minima as an insulating shell is dubious. For the lower boundary condition in the envelope, we set the radial current to vanish, consistent with the assumption of an insulating outer rocky layer.

We assume that all planets have a magnetic dipole field strength of $B = 1$ G arising from fluid motions within the iron-core, consistent with numerical simulations which find surface magnetic field strengths of super-Earths of $\sim 0.6 - 2$ G, a strength that varies little with mass (for $1 M_{\oplus} \le M \le 10 M_{\oplus}$) but increases with increasing iron-core mass fractions \citep{Driscoll_Olson}.  While it is possible that for close-in planets the stellar magnetic field can induce ohmic heating on the planet \citep{Laine_magstar_SE_2012 }, as a first step, we ignore these effects.  

We solve the induction equation following the prescription by \citet{Liu} and a numerical code based on a relaxation method to solve for the differential equations. We obtain the current in the radial and azimuthal directions, and the ohmic power dissipated at each radius and cummulatively. See Fig. \ref{induction} for the results of a typical calculation.

\section{Results}
We calculate more than 11,000 interior structure models spanning equilibrium temperatures between $1300$ and $1700$ K in intervals of $50$ K as well as 1325 K and 1375 K, masses of [2, 3, 4.5, 6, 9, 12, 15] $M_{\oplus}$ and envelope fractions of $1, 2, 3... 10\%$, and depositing ohmic heating rates of  $5\times10^{13}$ to $5\times 10^{18}$ W in logarithmic intervals of 0.1 until consistency is achieved. 

\subsection{Planet models without ohmic dissipation}

\begin{figure}
\includegraphics[width=0.45\textwidth,trim=0 0 0 0, clip]{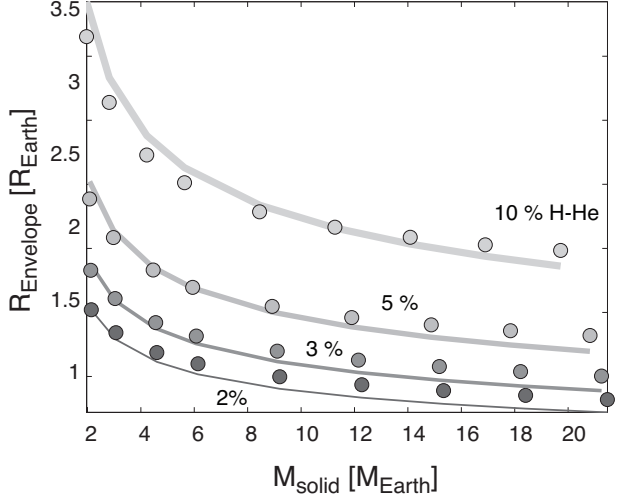}
\caption{The radius of the envelope for planets with various rocky core and envelope masses and no ohmic heating with $T_{eq}=1500$ K.}
\label{Renv_fit}
\end{figure}

We first obtain the thermal evolution and structure of planets without any additional heat sources.   One of the differences between mini-Neptunes and Jovian planets, is that the rocky interior, which is expected to contract negligibly in comparison to the envelope, makes up most of the planet's mass and hence greatly influences the overall size evolution of the planet. A fit to the envelope's size evolution is shown in Fig. \ref{Renv_fit} and can be summarized as:

\begin{multline}
R_{env} \equiv R - R_c \sim  1.4 R_{\oplus} \left(\frac{T_{eq}}{1600 K}\right)^{0.5}  \left(\frac{M_{core}}{10 M_{\oplus}}\right)^{-0.8} \\
\times  \left(\frac{M_{env}}{0.1 M_{\oplus}}\right)^{0.5}   \left(\frac{t}{\text{Gyr}}\right)^{-0.08}.
\end{multline}
\qr {The fit yields a mean error of 8\% and a maximum error of 64\%. This cooling law has a weaker dependence with time compared to the analytical prescription by 
\citet{Sari_1} who suggested a relationship for  Jupiter-like planets of $R(t) \propto t^{-0.25}$.  We attribute the differences mainly to the effects of the rocky core, which in the case of mini-Neptunes, dominates the structure.  We also find that the radius at the convection zone scales as $R_c \propto R^{0.7} M^{0.3} t^{-0.08}$.}

\subsection{Planet models with ohmic dissipation}

\qr{We then move onto calculating the ohmic dissipation in mini-Neptunes and the effect on their evolution. With the results on the internal structure, we calculate the conductivity profile and solve for the induction equation.  We take the cummulative ohmic power up to the radiative-convection zone as the available dissipation for inflation and feed this back into the interior structure and thermal evolution model as a constant heat source at the deepest point in the envelope (i.e. at $r = R_c$). We  iterate until the calculated cumulative ohmic dissipation at the $RBC$ matches the input at the 20 $\%$ level.}

In addition, for a subset of our models (with masses $M = [4.5, 9, 15] M_{\oplus}$, equilibrium temperatures $T_{eq} = [1300, 1500, 1700]$ K, and envelope fractions of $[2\%, 5\%, 10\%] $) instead of a constant heat source, we re-calculate the ohmic dissipation at the $RCB$ every 1 Myr by solving the induction equation, and feed this value back into the model. We find this subset of models incorporating the time-varying feedback to yield virtually identical final planet radii and atmospheric profiles as the models with constant coupling. In fact, we find the final equilibrium radius and atmospheric profile of a planet to be determined entirely by the amount of heat deposition regardless of history, consistent with the analytic models derived by \cite{Sari_1}. As a result, we adopt the constant ohmic dissipation prescription for the bulk of our models. 

\begin{figure}
\includegraphics[width=0.47\textwidth,trim=0 0 0 0, clip]{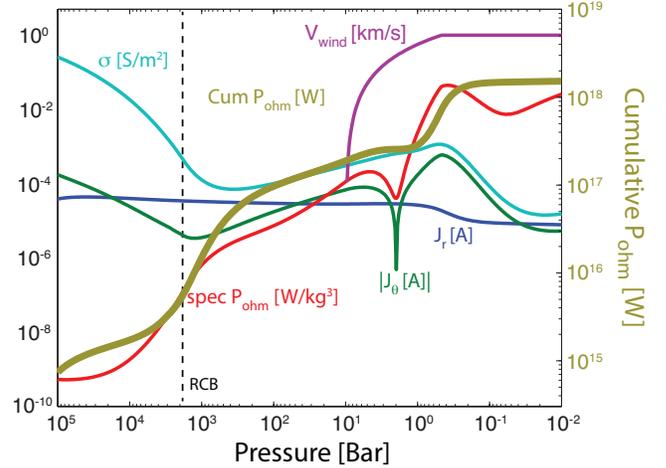}
\caption{Typical results to the induction equation. The induction equation is solved in a spherical 1D geometry given a wind profile (purple), and conductivity profile (cyan). The results are the radial current (blue), the azimuthal current (the absolute value is shown as green), and ohmic dissipation as a snapshot at 1Gy.  We compute the cumulative ohmic dissipation (olive green and right axis), and the ohmic dissipation per volume (red). The units of each parameter are shown in the figure and all but the cumulative ohmic dissipation correspond to the left (black) axis. This particular planet corresponds to a 9 $M_\oplus$- planet with 5\% H-He at 1500K. The radiative-convective boundary is shown as a dashed black line. }
\label{induction}
\end{figure}

\qr{A typical solution to the induction equation is shown in Fig. \ref{induction}.  This specific example is for a planet with $M=9M_\oplus$, 5\% H-He and $T_{eq}=1500$K. The wind profile (purple) is confined to the upper layers of the atmosphere and has a quadratic shape with a maxima of $1$ km/s.  The radial current (blue) varies mostly in the upper layers (above $\sim 1$ bar), and much less below it. The cumulative ohmic power (green olive with axis on the right)  follows closely the shape of the inverse of the conductivity (cyan) given that conductivity varies many orders of magnitude more than the radial or azimuthal currents (green). We also compute the ohmic power per unit mass (red). This specific ohmic power can be compared to the results \citet{BS11} showed in Fig. 4 for Jupiter-like planets. }

Similar to \citet{BS11, xuhuang, wu_lithwick}, we find that the largest ohmic values are obtained in the upper atmosphere where they are not efficient at inflating the planet. While the amount of ohmic dissipation within the convective zone is more than an order of magnitude smaller than in the upper atmosphere, it is still enough to halt the contraction of this planet during its evolution.

In a planet with exponential profiles for the conductivity, and a constant current $j$ across the interior of the planet, the ohmic power deposited below the radiative-convective boundary is equal to the power density at the boundary multiplied by the volume of a shell with thickness the conductivity scale height $H$:

\begin{equation}
P_{Ohm} \sim 4 \pi R_{RCB}^2 H_{RCB} j^2 \sigma_{RCB}^{-1}.
\end{equation}

Here, quantities with subscript $RCB$ indicate the parameter is evaluated at the radiative-convective boundary. In our numerical models, we find the best fit expression for the ohmic power to be:

\begin{multline}
\label{eq:pwr}
P_{ohm} \sim  2.6 \times 10^{16} W \left(\frac{R}{3R_{\oplus}}\right)^{2.2} \left(\frac{M}{10 M_{\oplus}}\right)^{-2.2} \\ \times \left(\frac{T_{eq}}{1500 K}\right)^{3.9}  \left(\frac{j}{10^{-4} A}\right)^{1.88} \left(\frac{\sigma}{10^{-3} \text{S/m}}\right)^{-1.2} .
\end{multline}

However, this fit has a RMS error or 41\% and a maximum error of 216\%  showing that a simple scaling fails to capture the complexity of the system. In fact, the fit does worse for larger planets with more envelope mass fractions, perhaps indicating a transition in the interplay of factors determining ohmic dissipation between mini-Neptunes and Jovian-like planets. 

\begin{figure}
\includegraphics[width=0.45\textwidth,trim=0 0 0 0, clip]{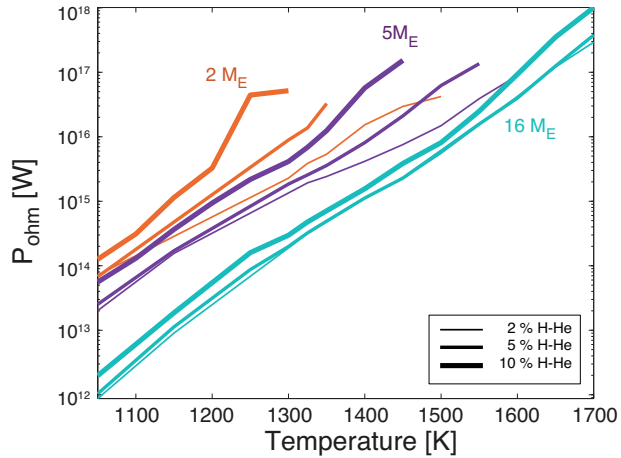}
\caption{Ohmic power generated within the convection zone as a function of equilibrium temperature for mini-Neptunes. We show the amount of ohmic dissipation below the radiative-convective boundary for planets of 3, 6 and 16 $M_\oplus$ and different envelope fractions spanning 2\% (thin lines), 5\% (medium lines) and 10\% (thick lines) H-He. Small, hotter planets generate more ohmic dissipation available for inflation.}
\label{OhmRCB_Teq}
\end{figure}

A useful result is to show the dependence of ohmic power on the equilibrium temperature of the planet for a variety of planet masses and compositions, which can be seen in  Fig. \ref{OhmRCB_Teq}. Given that the degree of partial ionization of metals increases exponentially with temperature, the relationship to ohmic power is very steep ( $P_{ohm} \propto T^{23}$).  We see that planets with smaller masses and larger gaseous mass fractions tend to experience more ohmic heating, since these planets have larger envelope radii and therefore a larger $R_{RCB}$.  This means that more ohmic power is generated in a hot small planet with a large gaseous envelope.  More importantly the amount of ohmic power generated in mini-Neptunes ranges between $10^{14}-10^{17}$, and is two to four orders of magnitude smaller than in exoJupiters (\citet{BS10, Perna_2010_hotJ} obtained ($P_{ohm}\sim 10^{19}$ for HD209458b).  

We now turn to estimating if this amount of ohmic power is enough to halt the contraction of mini-Neptunes.

\subsection{Time evolution of planets with ohmic dissipation}

\begin{figure}
\includegraphics[width=0.45\textwidth,trim=0 0 0 0, clip]{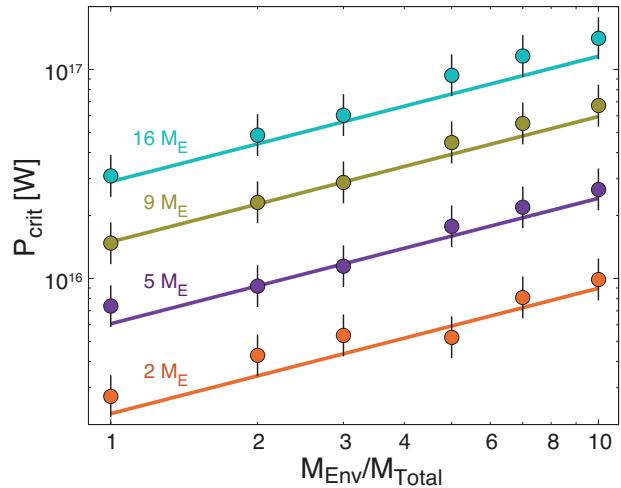}
\caption{Critical ohmic power needed to stop contraction at 1 Gy as a function of planet mass and envelope fraction. Planets with low mass and low volatile content require less energy for their contraction to halt. }
\label{Pcrit_Menv}
\end{figure}

\begin{figure}
\includegraphics[width=0.45\textwidth,trim=0 0 0 0, clip]{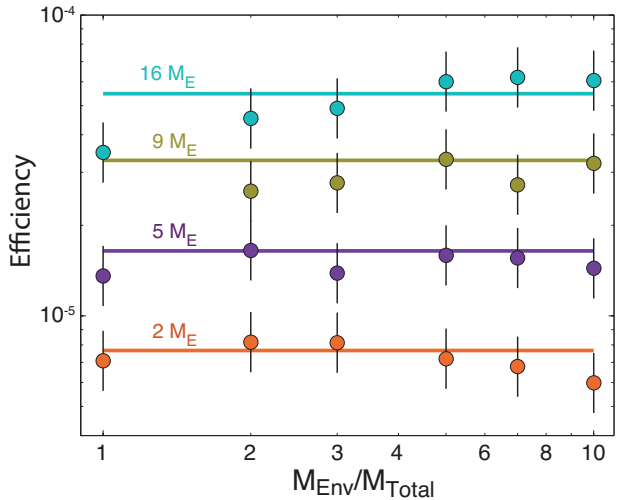}
\caption{Efficiency needed to stop the contraction at 1 Gy as a function of envelope fraction for different planet masses. Efficiency calculated as critical ohmic power over insolation energy ($\sigma T^4$).  }
\label{efficiency}
\end{figure}

When ohmic dissipation is introduced to a planet, we find that the main effect of such an energy source is to stop its contraction. Figure \ref{Revo_ohm} shows an example of the radius evolution of a $M = 16 M_{\oplus}$ planet with various levels of ohmic heating. We find empirically that, the amount of ohmic dissipation $P_{crit}$ required to halt the planet's cooling at $t = 1 $Gyr is given by (see Fig. \ref{Pcrit_Menv} and \ref{efficiency}):

\begin{equation} \label{eq:Pcrit}
P_{crit} \sim 1.9 \times 10^{16} W \left(\frac{M_{env}}{10 M_{\oplus}}\right)^{1.32}\left(\frac{f}{\%}\right)^{0.6}.
\end{equation}

Here, $M_{env} = M \times f$ is the total mass of the gaseous envelope. This fit has a mean error of 11\% and a maximum error of 48\%. Note that the choice of $t = 1 $Gyr is a stringent one, since it takes more energy to halt contraction at an earlier rather than at a later time. Compared to this nominal value of $10^{16}$ W,  the energy needed to halt HD209458b we calculate to be $3\times 10^{19}$ W or 3 orders of magnitude larger.

Notably, the amount of energy needed to stop the contraction of mini-Neptunes is of comparable magnitude or less than that which is generated through ohmic dissipation. One can compare Eqs. \ref{eq:pwr} and \ref{eq:Pcrit}, but preferably Figs. \ref{OhmRCB_Teq} and \ref{Pcrit_Menv} to arrive at this inference.

Thus, while the ohmic dissipation generated in mini-Neptunes is small (about 2 orders  of magnitude lower than typical values obtained for exo-Jupiter), it is enough to inflate mini-Neptunes (about 3 orders of magnitude lower than for exo-Jupiters). In short, mini-Neptunes can more easily be inflated with ohmic dissipation than their Jovian counterparts. 

We can also translate this critical ohmic power needed for halting contraction as an efficiency $\epsilon_{crit} \equiv P_{crit} / 4\pi R^2 \sigma T_{eq}^4$, defined as the total ohmic power divided by the amount of incident solar insolation (with zero albedo). We find empirically that

\begin{equation}
\epsilon_{crit} \sim 3 \times10^{-5} \left(\frac{M}{10 M_{\oplus}}\right)^{1.03} \left(\frac{T}{1500\text{K}}\right)^{-4}.
\end{equation}

This fit has a mean error of 14\% and a maximum error of 51\%. Interestingly, we observe that approximately, the critical ohmic efficiency does not depend on the gaseous fraction of the atmosphere. Note that this relationship is valid for mini-Neptunes planets and should not be used for Jovian planets that obey a different mass-radius relationship. We see that the ohmic efficiency required to inflate the atmospheres of mini-Neptunes is $\sim 10^{-5}$, orders of magnitude lower than the case of hot Jupiters, where values of $\sim 1 \%$ \citep{showman_guillot} are necessary to inflate their radii to account for observations.

\begin{figure}
\includegraphics[width=0.45\textwidth,trim=0 0 0 0, clip]{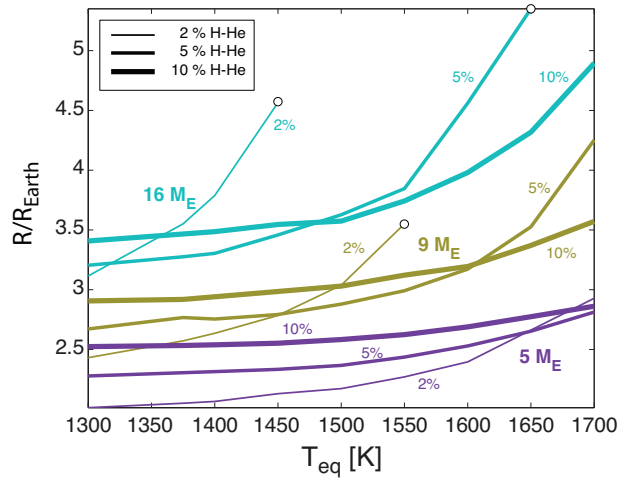}
\caption{The final evolved radius as a function of $T_{eq}$, when the effects of ohmic dissipation are taken into account. The planet mass is indicated in colour and the thickness of the line represents the amount of volatiles from 2, 5 and 10\% H-He.}
\label{R_Teq}
\end{figure}

Another useful way to see the effect of ohmic dissipation is to show the evolved radius of mini-Neptunes as a function of different equilibrium temperatures for different masses and envelope fractions.  Figure \ref{R_Teq} shows that the radius dramatically increases at equilibrium temperatures that depend  on the envelope mass fraction and planet mass and range between $1400 - 1600$K. 

For a given planetary mass at low temperatures (where ohmic dissipation is unimportant) the planets with more envelope fractions are larger, as expected. However, as the temperature and hence, ohmic dissipation increases, the trend gets overturned: the planets with less envelope fraction puff up to the point that they become larger than those with higher envelope mass fractions (compare lines with same color in Fig. \ref{R_Teq}). 

Counterintuitively, the $T_{eq}$ at which the radius dramatically increases happens at lower temperatures for more massive planets with a given envelope mass fraction (compare the lines with same thickness in Fig. \ref{R_Teq}).  

A corollary to these effects is that a hot mini-Neptune would be considerably puffier and susceptible to atmospheric evaporation than a cooler one, and could experience a runaway evaporation as it loses envelope mass in the early stages of evolution.  \blue{This fits nicely with recent work by \citep{Fulton:2017} that suggests there may be a deficit in occurrence rate distribution of planets with a radii between $1.5-2\, R_\oplus$  and that this gap seems to be at least partially shapped by stellar irradiation.}

We note that some planets with little H/He  ($\leq 2 \%$) can undergo ohmic heating so extreme that the planetary radius starts expanding ($dR/dt > 0$) instead of contracting during its thermal evolution (shown as circles in Fig. \ref{R_Teq}). We note that we do not explore this regime further.

For comparison to previous work, we obtain a simple scaling for the radius inflation with efficiency $\Delta R \propto \epsilon^{0.12}$,  and see that the dependence is shallower than for the case of hot Jupiters $\Delta R \propto \epsilon^{0.3}$ \citep{Sari_1}.

The fact that the radius of a mini-Neptune can be inflated due to ohmic dissipation means that when inferring the composition from mass-radius data, one can wrongly estimate the H/He content. It is clear from Fig. \ref{R_Teq} that there is a pervasive degeneracy in composition of hot mini-Neptunes. For example, a 16 $M_\oplus$ at $\sim 1360$ K and radius of $\sim 3.45 R_\oplus$ can be composed of 2\% or 10\% H-He (crossing of lines of same color). This degeneracy arises due to the fact that smaller H/He envelopes are more susceptible than the larger ones to inflation due to ohmic dissipation.  

\begin{figure}
\includegraphics[width=0.47\textwidth,trim=0 0 0 0, clip]{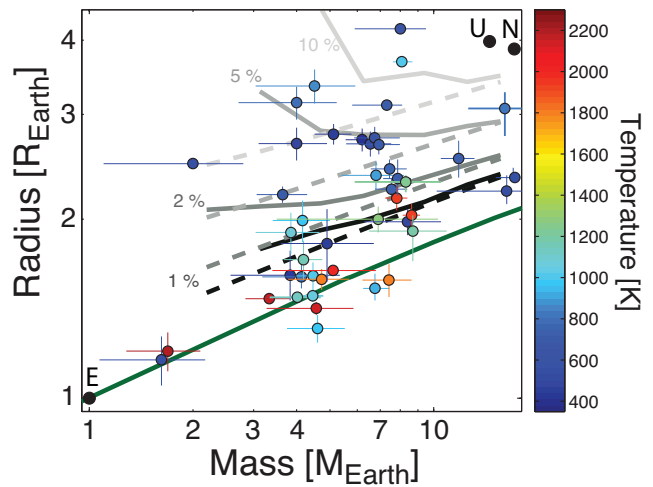}
\caption{Mass-radius relations for planets with various ratios of rocky material to gas. The dashed lines shows the mass-radius relations if no ohmic dissipation is present, while the solid lines incorporates ohmic dissipation for planets with an equilibrium temperature of 1400 K for four different envelope fractions: 1\%, 2\%, 5\% and 10\% H-He (ranging from dark to light grey).  Curves for the 5 and 10\% envelope fractions have been slightly smoothen out. Exoplanets with known masses and radii with error bars below 50\%  are shown and color coded by equilibrium temperature (calculated with an albedo of 0.3). Earth, Uranus and Neptune are shown for reference. The mass-radius relation for an earth-like composition is shown in green. }
\label{MR}
\end{figure}

Another way to look at this is with a mass-radius diagram. Figure \ref{MR} shows the effect of ohmic dissipation (solid lines) on the MR relationships compared to mini-Neptunes without it (dashed lines). It is clear that the most susceptible planets to experience radius inflation from ohmic dissipation are those with masses $\lesssim 8 M_{\oplus}$ and high envelope fractions $\gtrsim 5\%$ and $T_{eq} \ge 1400$K. 

\section{Discussion}

Our results show that, in the case of mini-Neptunes with close orbits around their host stars, ohmic dissipation in many cases can alter significantly the expected mass-radius relationship of such planets. The effect is more pronounced for less massive planets and planets with a smaller fraction of their mass in their gaseous envelope. For a $6 M_{\oplus}$ planet at $T_{eq} = 1500$ K with $10 \%$ gaseous content, the effects of ohmic dissipation can double the planet's radius, whereas the increase in radius is only $\sim 10\%$ for a $16 M_{\oplus}$ planet with the same composition. 

Like in previous studies of ohmic dissipation, we find that the amount of ohmic heating depends sensitively on temperature, which plays two roles in ohmic dissipation. First, the amount of partial ionization in the wind zone depends sensitively on the temperature there, and as a result the induced magnetic field has a sharp exponential dependence on the temperature. At the same time, the ohmic heating is inversely proportional to the conductivity in the convection zone, and increasing the temperature there leads to a decrease in ohmic heating. The net effect of varying $T_{eq}$ is dominated by the first factor, and as a result we see a steep dependence of $P_{ohm}$ on $T_{eq}$ with ohmic dissipation becoming increasingly relevant for $T_{eq} \gsim 1400$ K, similar to what \citet{BS11} observed for hot Jupiters.

This scaling result agrees qualitatively with \citet{xuhuang}, who found that the ohmic power increases for low-mass planets with extended radii. For a fixed temperature of the isothermal atmosphere $T_{iso}$ and the induced magnetic field $B_{\phi0}$, \citet{xuhuang} propose a scaling of $P_{ohm} \propto R^{2.4} /M$. In comparison (and fixing the same quantities) we obtain $P_{ohm} \propto R^{3.3} M^{-0.8}$. 
It is easier to puff up lower-mass planets for two reasons: for one, the lower gravity leads to a greater scale height, increasing the volume in which ohmic dissipation takes place and thereby the ohmic dissipation; moreover, the amount of dissipation necessary to halt a planet's contraction $P_{crit}$ increases with mass, and therefore less massive planets require less power to remain puffed-up.  \qr{The differences in the exponents of the scalings may arise from several sources. First, \citet{xuhuang} proposed their scalings based on Jovian planets and not planets with a large core fraction and incipient envelopes. Second, our choice of placing all the ohmic dissipation at the bottom of the envelope as a constant source is an upper bound to how much inflation can happen. And lastly, we do not take into account the Lorentz drag and have chosen a different set of boundary conditions. }

We now turn to discussing these assumptions. For simplicity, we account for ohmic dissipation by placing it as a shell source of luminosity at the base of the planet's gaseous envelope, ignoring its evolution with time and its radial profile. By depositing all ohmic heating deep in the planet's envelope our results constitute an upper limit to the amount of dissipation and inflation. For example, \citeauthor{xuhuang} found that ohmic heating tends to push the convection zone inwards, decreasing the amount of ohmic dissipation inside the convection zone.  A precise determination of the location of the radiative-convective boundary is crucial for accurate measurement of the ohmic flux. 

On the other hand, the second approximation turns out to be an excellent one. This is due to two reasons. First, the eventual configuration of the planet appears to depend solely on the final heat deposition at equilibrium; taking into account the time evolution of ohmic heating only serves to delay the planet's contraction as ohmic dissipation generally grows weaker with time, without implications on the final planet radius.

Secondly, the delay itself tends to be small, as the ohmic efficiency tends to evolve slowly. Across all our models, we found that after the initial $\sim$Myr rapid cooling phase, the amount of ohmic dissipation only decreases by $56 \%$ on average from $t = 10$ Myr to $t = 10$ Gyr. Comparisons between our model with evolving feedback and the constant dissipation case show negligible differences in both time evolution and final planet radius.

Another approach we take that is different in comparison to previous studies is our choice of boundary conditions for solving the induction equation. Our nominal choices are to assume that the radial current $j_r$ vanishes at an inner and outer boundary, forming a closed current loop. We take the transition from rocky interior to gaseous envelope to be our inner boundary. This boundary condition assumes that the rocky outer shell is an insulator. For the outer boundary condition, we take the radial current $j_r$ to vanish at a set pressure of $P = 10^{-6}$ bars.

However, we explore two different boundary conditions and assess the effects on ohmic dissipation. For the bottom boundary condition, we also explore the condition proposed by \citet{Liu} where $j_\theta - j_r/r = 0$ for Jupiter at $r=0$.  This choice of boundary condition had a minor effect, with lower values of cumulative ohmic dissipation by about $20\%$. 

For the top boundary condition, we explore the one suggested by \citet{BS10}.  They found that the atmospheres of hot Jupiters feature a local temperature minimum, which motivated them to set the radial current to vanish at this shell. This boundary condition turned out to match ours well for $T_{eq} \lsim 1400$ K; at higher equilibrium temperatures, the conductivity at the temperature minima is not sufficiently insulating to enable \citet{BS10} boundary condition. In comparing both treatments, our choice of boundary condition (radial current vanishing at one millibar) yields an ohmic dissipation that is $\sim 50 - 100$ times higher than if we set the insulation at the conductivity minima.  \qr{This points to the importance of the boundary conditions in affecting the results.  We admit that our boundary conditions are simple and based on the assumption that the currents loop in the interior so that at some point in the atmosphere and the bottom of the envelope the radial currents have to vanish. However, we have not investigated other more complicated boundary conditions such as open currents and complicated geometries.}

Our assumption that the rocky outer shell is insulating enough so that the currents originating in the winds do not penetrate the rocky interior stems from our goal of invoking a simple treatment for mini-Neptunes in light of the many uncertainties about their interior. We know little about the possible phase structure of super-Earths and mini-Neptunes.  Perhaps they do not have a well segreated iron core from an overlying silicate mantle, which would put into question how and where would a magnetic field may be generated.  Or even under this mixed structure,  there could be regions of high conductivity (semi-conductive state of mixture, presence of melts, favorable spin states, etc) that can influence the type of magnetic field being generated.  Or, perhaps the electrical currents generated in the wind can penetrate the planet throughout because at the relevant temperatures and pressures conduction is high enough and phase transitions do not hinder the induced currents.  We leave these scenarios for future work and focus on the simple treatment where the currents are confined to the envelope of the mini-Neptunes. 

As previously mentioned, we do not include Lorentz drag that is expected to reduce the ohmic power at high enough temperatures at this point.  Because we focus on planets with equilibrium temperatures of up to 1700 K, we expect Lorentz drag to yet have become dominant to affect the flow and hence, our results.

\section{Summary}

In summary, we investigate ohmic heating in planets with a rocky interior surrounded by an envelope of hydrogen and helium gas, based on a simplified model in which the evolution of the atmospheric structure and \qr{the ohmic dissipation are calculated separately and coupled via iteration}. 

\qr{We find that ohmic dissipation increases with shrinking planetary mass and increasing envelope mass fraction. The overall normalization depends on the exact modelling details.  In addition, the energy needed to halt contraction during the thermal evolution increases with planetary mass and envelope fraction. Thus, in combination, the ohmic dissipation available is enough to inflate the radius of low-mass mini-Neptunes. }

\qr{The implications are two-fold. First, it means that there is an added degeneracy to the problem of inferring the composition of mini-Neptunes.  There is a trade-off between H/He content and ohmic dissipation.  In fact, even while correcting for ohmic dissipation, planets with the same mass, radius and equilibrium temperature can have different H/He contents due to the fact that incipient envelopes are much more susceptible to inflation than more massive envelopes. For planets with equilibrium temperatures in excess of $1400$ K, one would first have to correct the radius for inflation caused by ohmic dissipation and calculate the possible ensembles that yield the radius for a given mass. Without this treatment, the H/He inferred could be wrong.  } 

\qr{Second, ohmic dissipation may be a very important effect when considering evaporating atmospheres of highly irradiated planets.  Already they are hot to drive evaporation, so that ohmic dissipation may be expected to be present, making the planets puffier than otherwise and susceptible to mass loss.   This might explain why for the planets with measured masses and radii, there is a lack of hot mini-Neptunes compared to warm ones.}

\acknowledgements{}

We would like to thank Mathieu Havel for multiple discussions that
helped benchmark our results, Konstantin Batygin for comments on the first draft of this paper, and the anonymous reviewer for her/his insights.   This work was funded under grant NSERC RGPIN-2014-06567.

\bibliography{Notes}
\bibliographystyle{apj}

\end{document}